\documentclass[amsmath,amssymb,nofootinbib,prd,10pt,floatfix]{revtex4}
\usepackage{bm}
\usepackage{amssymb}
\usepackage{graphicx}
\usepackage{epstopdf}

\begin{document}

\title{Orbital evolution for extreme mass-ratio binaries: conservative self forces}

\author{Lior M.~Burko}

\address{Department of Physics, University of Alabama in Huntsville, 
Huntsville, Alabama 35899}

\begin{abstract}

The conservative dephasing effects of gravitational self forces for extreme mass-ratio 
inspirals are studied. Both secular and non-secular conservative effects may have a 
significant effect on LISA waveforms that is independent of the mass ratio of the 
system. Such effects need to be included in generated waveforms to allow for accurate 
gravitational wave astronomy that requires integration times as long as a year. 

\end{abstract}


\maketitle


A compact object of mass $\mu$ (such as a neutron star or a stellar mass black hole) 
that spirals into a supermassive central object of mass $M\gg\mu$ (such as a black 
hole in the center of a galaxy) experiences, in 
addition to dissipative radiation reaction, also conservative effects that are caused 
by the object's self force \cite{poisson-review}. Dissipative effects determine 
the {\em adiabatic} waveforms to $O(\mu/M)^{-1}$ \cite{burko-I}. Conservative effects 
determine corrections to the waveforms at $O(\mu /M)^{0}$, i.e., they are independent of 
the mass ratio. The latter 
include both secular and non-secular effects. Secular conservative effects include 
an additional precession of the periastron that results in a conservative dephasing 
of the gravitational waveforms \cite{burko-ijmp,poisson}. Indeed, the dephasing of the 
gravitational waveform per revolution is at $O(\mu /M)$ \cite{burko-ijmp}, but as the 
total number of orbits is at $O(\mu /M)^{-1}$, the cumulative dephasing effect is 
independent of the mass ratio. Non-secular effects 
include a change to the frequency of the orbit, that is not accompanied by a change to 
the object's energy, angular momentum, or Carter's constant \cite{burko-cqg}. Again, this 
effect results in dephasing at $O(\mu /M)^{0}$, which comes about because non-secular 
conservative effects are smaller than dissipative effects by $O(\mu /M)$ 
\cite{burko-I,drasco}.  

In this Paper we estimate the magnitude of such effects for EMRI's in the 
post-Newtonian framework. The conservative piece of the particle's self force is 
an additional force that acts on the particle. Such a force results in an additional 
precession of the periastron. At the 1PN level, this additional 
rate of change of the periastron advance 
$\langle{\dot\omega}\rangle\,\sim \mu M^{1/2} r^{-5/2}$. 
Throughout this Paper we use geometrized units, 
in which $G=1=c$. We denote the semi-major axis with $r$, and the eccentricity with $e$. 
As the radiation reaction time 
scale is $\tau\sim r/{\dot r}\sim \mu^{-1}M^{-2}r^4$, the conservative dephasing of the 
gravitational waveform scales with 
$\langle{\dot\omega}\rangle\,\tau\sim (r/M)^{3/2}\sim v^{-3}$, where 
$v$ is the particle's velocity \cite{poisson-capra}. 
We show below that this crude approximation 
captures well the order of magnitude of the phenomenon, specifically for low 
eccentricities.  

Here, we seek a crude order-of-magnitude estimate of the conservative dephasing 
effect using dissipative terms that are accurate to to 3.5PN, and conservative 
terms to 2PN\footnote{Notice, that Eqs.~(27)--(31) in Ref.~\cite{barack-cutler} are 
given to leading order in $\mu$. The conservative self force effect requires an 
expansion to one order higher in $\mu$ in Eq.~(29).} \cite{barack-cutler} for 
an object $\mu$ in bound motion around a Schwarzschild black hole of mass $M$. Note, 
that some of the technique needed for generalization to a Kerr black hole is 
already available \cite{burko-cqgI}. Specifically, we take
\begin{equation}
{\dot r}=-\frac{32}{5\pi}\frac{\mu}{M^3}\frac{r}{f}\frac{(2\pi Mf)^{11/3}}
{(1-e^2)^{9/2}}\times [\alpha (1-e^2)+\beta (2\pi Mf)^{2/3}]
\end{equation}
where $f=(M+\mu)^{1/2}/(2\pi r^{3/2})$ is the orbital frequency, 
$\alpha=1+(73/24)e^2+(37/96)e^4$ and $\beta=1273/336-(2561/224)e^2-(3885/128)e^4-
(13147/5376)e^6$. The periastron advance is given by 
\begin{equation}
\langle{\dot \omega}\rangle=6\pi\frac{M}{r}f(1-e^2)^{-1}\left[ 1+\frac{M}{r}\frac{26-15e^2}
{4(1-e^2)}+\frac{\mu}{M}+\frac{\mu}{M}\frac{M}{r}\frac{(26-15e^2)}{2(1-e^2)}
\right]\, .
\end{equation}
Here, the terms proportional to $\mu$ are the effect of the Newtonian and 1PN self force, 
respectively. The former is, in fact, a consequence of the redefinition of the total mass 
of the system, $M+\mu$ \cite{poisson-detweiler}. The rate of change of the eccentricity 
$e$ is given by \cite{junker-schafer}
\begin{equation}
{\dot e}=-\frac{1}{15}\frac{\mu M^2}{r^4}\frac{e}{(1-e^2)^{7/2}}\left[(304+121e^2)
(1-e^2)-\frac{1}{56}\frac{M}{r}\left(\gamma+\frac{\mu}{M}\delta\right)\right]\, 
\end{equation}
where $\gamma=133640+108984e^2-25211e^4$ and $\delta=4(9352+8421e^2+847e^4)$.

\begin{figure}
\includegraphics[width=12.0cm]{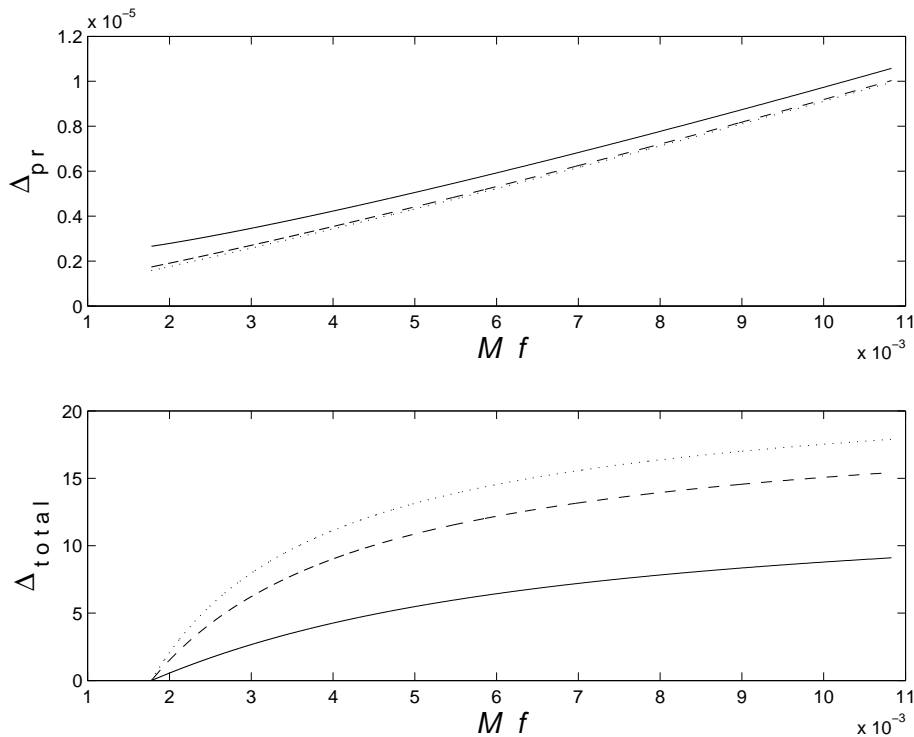}
\caption{The conservative dephasing due to periastron precession as a 
function of the orbital frequency. Upper panel: The dephasing per revolution 
$\Delta_{\rm pr}$. Lower panel: The total cumulative dephasing $\Delta_{\rm total}$. 
In all cases the integration starts at semi-major axis of $r/M=20$, and the initial 
eccentricities are $0.1$ (dotted curves),  $0.3$ (dashed curves), and $0.6$ (solid 
curves).}
\label{dephase1}
\end{figure}

Next, we integrate these equations for a quasi-elliptic orbit, and find the periastron 
advance. Figure \ref{dephase1} shows the additional 
periastron advance per revolution and the 
cumulative additional 
periastron advance as functions of the orbital frequency, for three values of 
the initial eccentricity $e_i$. We do not show in Fig.~\ref{dephase1} the periastron 
advance due to geodesic motion. 
The dependence of the total conservative dephasing 
on $e_i$ is shown in Fig.~2. Clearly, in the post-Newtonian framework, 
the secular conservative dephasing is appreciable, and may be an important effect for LISA. 


\begin{figure} \label{dephase2}      
  \includegraphics[height=.4\textheight]{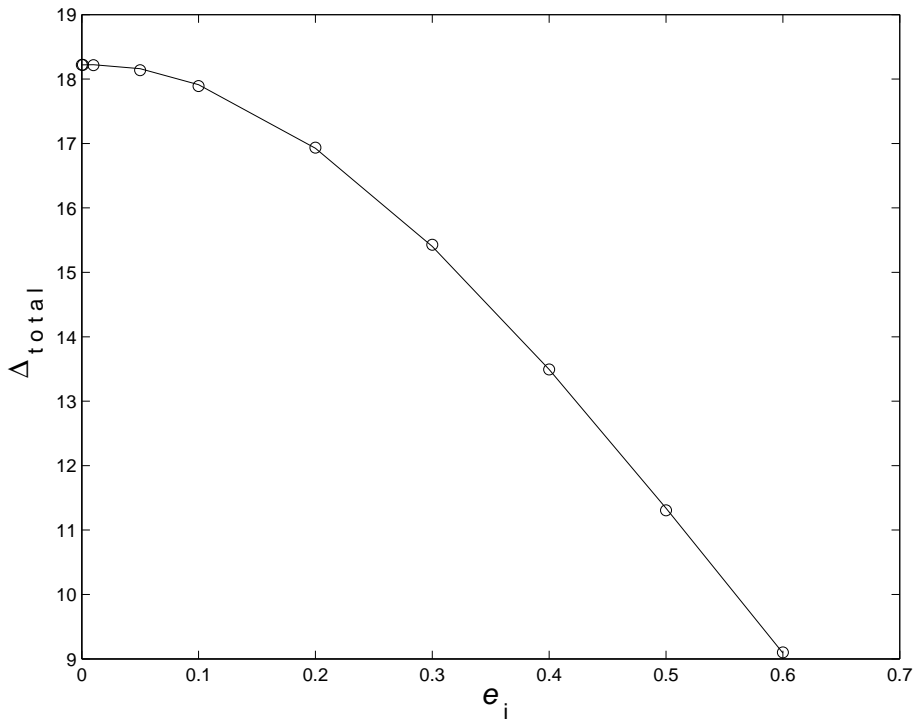}
  \caption{The cumulative conservative dephasing $\Delta_{\rm total}$ as a function 
of the initial eccentricity $e_i$. In all cases integration starts at semi-major 
axis of $r/M=20$. The circles depict the numerical data, and the solid curve is the 
fitted curve of $18.219+0.833e_i-41.4e_i^2+24.4e_i^3$.}
\end{figure}

Notice, that while the method of \cite{burko-ijmp} was correct, the assumed radial 
self forces that was used there was unrealistic. In fact, Ref.~\cite{burko-ijmp} 
assumed a radial self force at the 3PN order. In fact, the self force enters already 
at the 1PN level for gravitational self force (unlike scalar field self forces, for 
which the radial self force indeed is a 3PN effect \cite{burko-prl}).

The non-secular conservative effects were studied in \cite{burko-cqg} for the special 
case of quasi-circular orbits. Their effect on the orbital frequency is given by
\begin{eqnarray}
\frac{\,\Delta f}{f_0}&=&\frac{1}{2}\frac{\mu}{M}\left\{\left[1-2(2\pi Mf_0)^{2/3}
+\frac{61}{4}(2\pi Mf_0)^{4/3}+\cdots\right]\right. \nonumber \\
 &-&\left. 3\frac{\mu}{M}(2\pi Mf_0)^{2/3}
\left[1-\frac{65}{12}(2\pi Mf_0)^{2/3}+\cdots\right]+\cdots\right\}\, ,
\end{eqnarray}
where $f_0=M^{1/2}/(2\pi r^{3/2})$. This shift in frequency 
is caused by a conservative radial self force, that can be found using the following 
method: Consider the correction to Kepler's third 
law, in a post Newtonian expansion:
\begin{eqnarray}\label{omega}
\Omega^2 &=& \frac{M+\mu}{R^3} \left[ 1+(-3+\nu)\gamma + 
\left(6+\frac{41}{4}\nu+\nu^2\right)\gamma^2 \right.
\nonumber \\
&+& \left.
\left(-10+\rho\nu+\frac{19}{2}\nu^2+\nu^3\right)\gamma^3 +
\cdots \right]\, ,
\end{eqnarray}
where $R$ is the harmonic radial coordinate, $\nu:=M\mu/(M+\mu)^2$,  
 $\gamma :=(M+\mu)/R$,
and $\rho$ is a certain (known)  expression
\cite{blanchet-faye}.
Note, that  while $\Omega^2$ is a gauge invariant
quantity, it is here expressed in terms of gauge dependent quantities,
specifically the harmonic coordinate $R$.
We first recognize that the $\nu$-independent terms inside the square brackets in 
Eq.~(\ref{omega}) are the
leading terms in the expansion of $(1+\gamma)^{-3}$ for $\gamma\ll 1$. Using this, 
we notice that
$R^{-3}(1+\gamma)^{-3}=r^{-3}[1-3\mu/r+O(\mu^2)]$. With this substitution, we express 
the angular velocity as
\begin{eqnarray}\label{omega_sch}
\Omega^2 = \frac{M}{r^3} &+& \frac{\mu}{r^3} - 2\,\frac{M\mu}{r^4} + \frac{61}{4}
\frac{M^2\mu}{r^5}+\cdots \nonumber \\   
&-&3\frac{\mu^2}{r^4}+\frac{65}{4}\frac{M\mu^2}{r^5}+\cdots\, ,
\end{eqnarray}
where following the Keplerian term on the rhs we present in the first line of 
(\ref{omega_sch})
the Newtonian self force correction \cite{poisson-detweiler}, followed by the 1PN 
and 2PN first-order 
self-force corrections,
and where in the second line we present the Newtonian and 1PN second-order self-force 
corrections.

From Eq.~(\ref{omega_sch}) we can extract the form of the radial self force, in the 
post Newtonian gauge \cite{burko-cqg}:
\begin{eqnarray}\label{fr}
f_r &=& -\frac{\mu^2}{r^2} - \frac{\mu^2M}{r^3}-\frac{73}{4}\frac{\mu^2M^2}{r^4}
+O(\mu^2M^3/r^5)\nonumber \\
&+&2\frac{\mu^3}{r^3}-\frac{37}{4}\frac{\mu^3M}{r^4}+O(\mu^3M^2/r^5)\, .
\end{eqnarray}
In the first line of Eq.~(\ref{fr}) we present the first order self force, and 
in the next line the second order self force.

Figure 3 shows the waveforms for an orbit that starts at $r=10M$, 
and decays to the inner-most stable circular orbit (ISCO) at $r=6M$
for self-force correction at the Newtonian, 1PN and 2PN orders. Not surprisingly,
close to the ISCO the post Newtonian approximation breaks down. 

\begin{figure} \label{figure3}
  \includegraphics[height=.4\textheight]{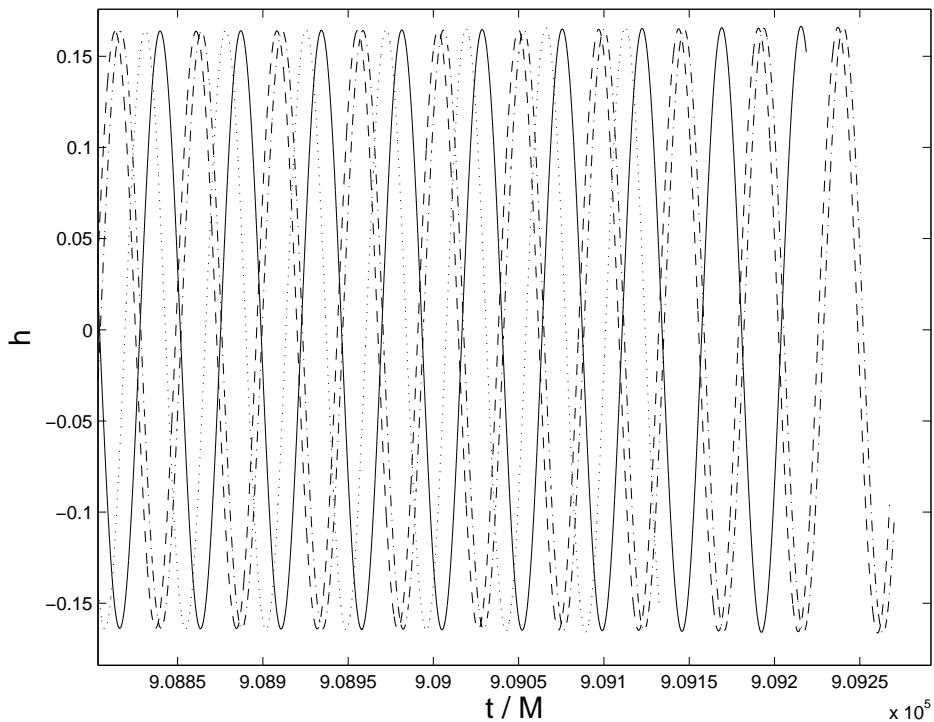}
  \caption{The waveform for an orbit starting at $r=10M$ at $t=0$, and decaying to the 
ISCO at $6M$ for $\mu=10^{-4}M$. Dotted line: no self force. Dash--dotted line: 
Newtonian self force. Dashed line: 1PN self force. Solid line: 2PN self force.
}
\end{figure}
Figure 3 suggests that this non-secular dephasing effect may be large enough 
to affect LISA observations.


\section*{Acknowledgments}
The author wishes to thank Leor Barack and Eric Poisson for discussions. This work 
was supported in part by NASA EPSCoR grant No.~NCC5--580.

\end{document}